# To Use or Not to Use: Graphics Processing Units (GPUs) for Pattern Matching Algorithms

D. R. V. L. B. Thambawita
Department of Computer Science and Technology
Uva Wellassa University
Badulla, Sri Lanka
vlbthambawita@gmail.com

Roshan Ragel and Dhammika Elkaduwe
Department of Computer Engineering
University of Peradeniya
Peradeniya, Sri Lanka
[ragelrg,dhammika.elkaduwe]@gmail.com

*Abstract*— String matching is an important part in today's computer applications and Aho-Corasick algorithm is one of the main string matching algorithms used to accomplish this. This paper discusses that when can the GPUs be used for string matching applications using the Aho-Corasick algorithm as a benchmark. We have to identify the best unit to run our string matching algorithm according to the performance of our devices and the applications. Sometimes CPU gives better performance than GPU and sometimes GPU gives better performance than CPU. Therefore, identifying this critical point is significant task for researchers who are using GPUs to improve the performance of their string matching applications based on string matching algorithms.

## I. Introduction

String matching algorithms are used to identify a specific string from a large database or a file. Several kinds of string matching algorithms are available targeting various kinds of applications. In brief, string matching algorithms are heavily used in applications such as bioinformatics, cryptography, anti-virus software and other specific applications that do patterns matching. A frequent problem in this area is the high computational power demands of executing string matching algorithms [1]. Researchers have turned to the concept of parallel computing to overcome these difficulties.

However, the traditional CPU (Central Processing Unit) based approaches might not be the best solution for performing parallel computing as their cost and scalability issues. GPU (Graphical Processing Unit) is so far one of the best cost effective solutions in this era to face this problem with massively parallel computing technology [2]. Present NVIDIA GPUs contain hundreds of cores that are capable of running thousands of light weight threads as opposed to a few numbers of threads available in current CPU [3]. Architecture difference between CPU and GPU can be identified simply by looking at Fig. 1 [4]. A normal CPU has several ALUs (Arithmetic Logic Units), a control unit to control those ALUs, fast cache memory and DRAM (Dynamic Random Access Memory) when consider with GPU which has hundreds of ALUs, several control units, several cache memories and one DRAM. These hundreds of cores are grouped into several multiprocessors for handling easily.

To harness the power of GPUs, NVIDIA has embodied a parallel computing platform for developers who are targeting parallel computing. Growth of the computational power of NVIDIA GPUs can be identified by looking at the Fig 2 [4]. It's X axis shows years of technology introduced and Y axis shows theoretical throughput of corresponding technology. Through this graph it is easy to identify that GPU has better through than CPUs.

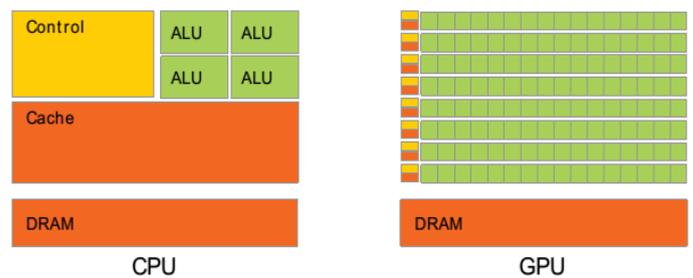

Fig. 1. CPU vs GPU Architecture

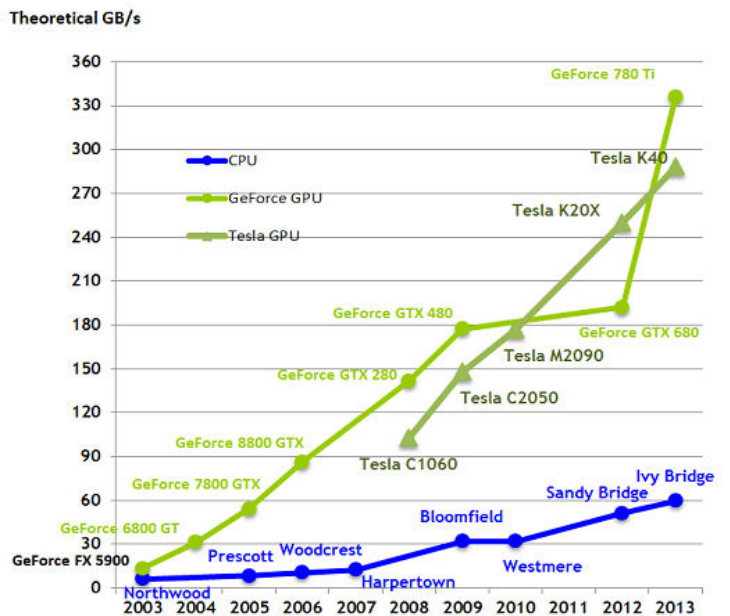

Fig. 2. GPU vs CPU Performance

In this paper, the performance of a string matching algorithm known as Aho-Corasick is compared on a CPU and a GPU in order to identify the suitable device during different situation for better performance. The GPU implementation (parallel version of the Aho-Corasick algorithm) has been done using CUDA (Compute Unified Device Architecture) [5-6] and executed on a state of the art Graphic Processing Unit.

Although, GPUs can be used for parallel computing, they cannot be used in every situation that need parallelism, as it may harm the performance of applications depends on the workload. Therefore, the performance comparison between CPU and GPU has to be done properly [7] to identify the suitable device based on the workload. Therefore, we analyse this critical point by using the CPU and GPU versions of the Aho-Corasick algorithm with changing input data size as the major parameter. In addition, we also considered the initiating time of the GPU and the effect of idle processes of the OS (Operating System). In this experiment, serial algorithm of Aho-Corasick has been used on CPU and parallel version of the algorithm has been used on GPU as shown in Fig. 3. Here the CPU processes character by character (C) of input stream serially and GPU processes all characters of the input parallelly to give outputs (N).

In this experiment, a serial version of the Aho-Corasick algorithm that runs on a CPU was used with a parallel version that runs on a GPU to compare performance of these two devices. However, parallel version of the Aho-Corasick algorithm that runs on CPU was not used in this initial stage of the experiment because the CPU serial version is the worst case on the CPU side. This CPU serial version is considered to prove that it can also serve better in some situations. Therefore, the parallel version of Aho-Corasick algorithm on CPU should have better performance than the serial version of the algorithm on CPU [8].

For this experiment, we used a serial version of Aho-Corasick algorithm called Multifast and a parallel version of Aho-Corasick algorithm called PFAC (Parallel Failure less Aho-Corasick). Then those two algorithms were run on CPU and GPU for analysing to use or not to use GPU for pattern matching algorithm.

## II. AHO-CORASICK ALGORITHM

The Aho-Corasick algorithm is a simple and efficient algorithm to locate all occurrences of any of a finite number of keywords in a string text [9-10]. This algorithm consist of a finite state pattern matching machine constructed by keywords and a process to search a given text string in a single pass. In this process, there are three main functions called Goto, Failure and Output. The Goto function identifies next location of the pattern matching machine of Aho-Corasick algorithm in order to identify correct state according to the input character stream. The Failure function identifies failure node when there is no output from Goto function that matches according to the input character stream. The Output function gives outputs of matched patterns of current state of the pattern matching machine [11-12].

In this experiment, we used two type of Aho-Corasick algorithms called "multifast" and "PFAC"[13-14]. The multifast is general Aho-Corasick algorithm with failure transitions and without any parallel processing techniques. A sample pattern matching machine used for Multifast is shown in Fig. 4. A link list is the basic data structure used to construct pattern matching machine in the multifast algorithm. Using these link list data structure, the Goto function and the Failure function have been defined. However PFAC uses another data structure with some matrix implementation to remove link list from the algorithm because implementing link list on GPU is a significant task and early GPUs gave low performance when using link list. One of the biggest problems with PFAC is getting the sub patterns of the large patterns as the output [15-16]. That is, although, the original Aho-Corasick algorithm outputs sub patterns, PFAC implementation ignored this feature, mainly due to the limitation they had in the implementation (not using linked data structure).

However, the PFAC implementation is another version of Aho-Corasick algorithm with parallel processing capability on GPU without failure transitions to get better performance rather than getting complete output like the original CPU version of Aho-Corasick algorithm [12].

The pattern matching machine without any failure links is used parallelly in PFAC to achieve a better performance using high end GPUs [17-18]. A sample failure less pattern matching machine data structure is shown in Fig. 5.

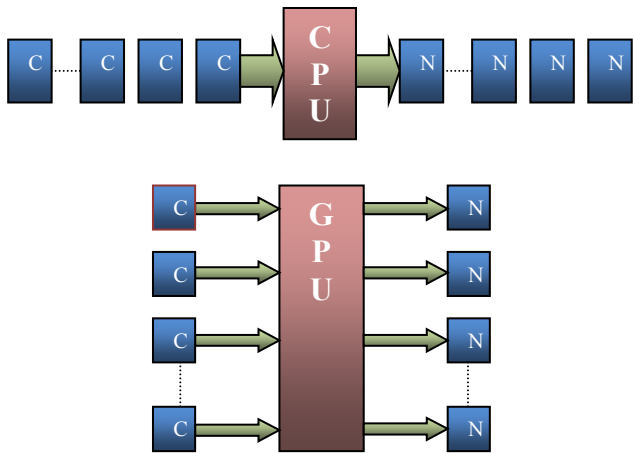

Fig. 3. Serial Processing Vs Parallel Processing

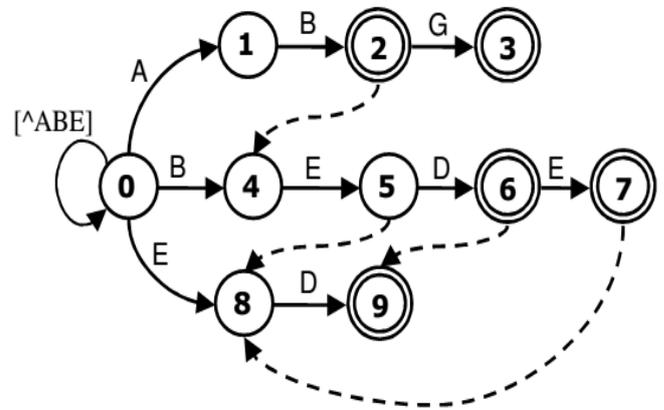

Fig. 4. Aho-Corasick pattern matching machine for "AB", "ABG", "BEDE", and "ED" with failure links

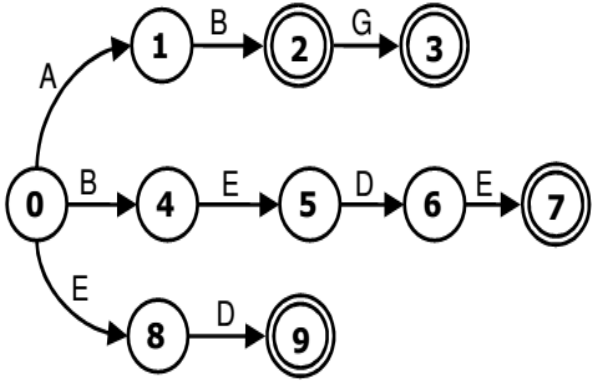

Fig. 5. Failure-less-AC pattern matching machine to find "AB", "ABG" "BEDE", and "ED"

III. EXPERIMENTS

The multifast and PFAC were run on CPU (Intel Core i5-3230M CPU @ 2.60GHz 2.60GHz) and GPU (NVIDIA GeForce GT 740M) respectively. As the patterns we used five (5) permutations using twenty six (26) characters of English alphabet (Ex: abcdefghijklmnopqrstuvwxyz).

We added English alphabet repeatedly into the text file to be used as input text file for the experiment. Then, the size of the input text file was increased gradually by adding initial text load repeatedly. For both multifast and PFAC implementations, the same input load was used at a time to identify the critical points when we were changing the text file load.

When measuring the time in this experiment, we considered few important facts like GPU initiating time and time used for giving output to the user using the printf() function because these facts directly affected the time measured. The GPU initiating time was ignored in time measured, because this time is considerably very large and this time is taken only in first time use of GPU. Therefore, we started time measurement after calling one of the available CUDA function like cudaFree(). The printf() functions were also removed from the implementations because printf() function is common factor to both multifast and PFACE to get outputs.

In addition to above facts, OS also plays a major role in the experiments. Ubuntu 12.04 LTS (x64) version has been used in our experiment. However hybrid GPUs (using two GPUs like Intel and NVIDIA in the same time) are not compatible with Ubuntu OS. Therefore, when the test program is run, the GPU had some idle processes related to the graphic processing part of our machine. However, we used "init" mode of the Ubuntu by stopping "gdm" services to overcome this problem and got the result with reduced interruptions from idle processing part of the GPU.

IV. RESULTS AND DISCUSSION

In the first step, we changed the file size from 16375 bytes to 671375 bytes to identify that what is the crossing point of time taken by CPU vs. GPU. The graph in Fig. 6 was plotted for the first experiment with the variation mentioned.

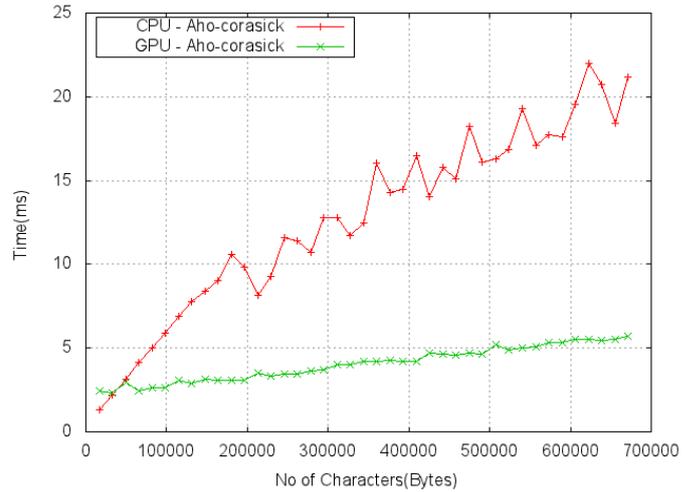

Fig. 6. Aho-Corasick with CPU and GPU

One major changing point can be identified on the above graph (Fig. 6) is closer to 50,000 bytes. Another experiment has been done to clarify the above critical point by changing the file size from 1310 to 53710 bytes and the graph in Fig. 7 was obtained.

The changing point of time taken by CPU and GPU can be identified clearly using the above graph (Fig. 7) and it is around 40,000 byte (40kB). Some disturbances can be seen on the above graph and it has happened due to the idle processes of the CPU and GPU. However, removing these idle processes is difficult with a personal computer.

According to the above result, we can see that the GPU is not an option for all the time to get the best performance in string matching algorithms like Aho-Corasick. Therefore, we have to think before using GPU for our applications. If we consider an application with string matching algorithm like Aho-Corasick and input file size is less than the critical point mentioned above then using GPU will give less performance than using CPU. Therefore, identifying this important point should be the major pre-experimental step before publishing your GPU parallelism based applications.

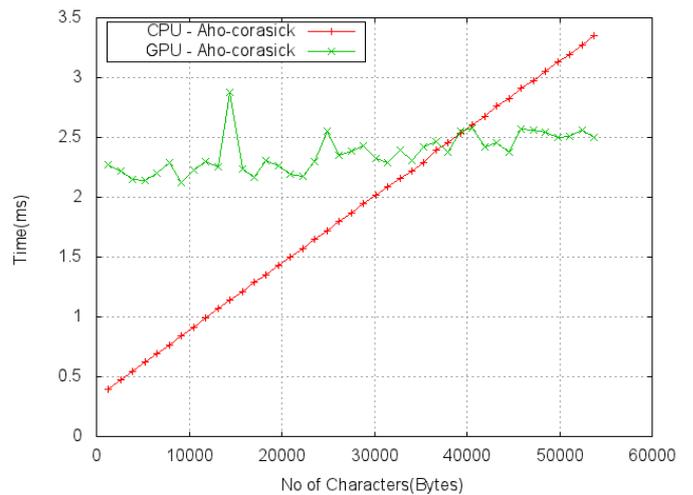

Fig. 7. Aho-Corasick with CPU and GPU (reduced load)

What are the facts directly affecting this point of changing the performance of CPU and GPU? The results in Fig. 8 were obtained to identify the one of the facts affecting this point by changing number of pattern from 5 to 10.

When data load of the algorithm goes higher, the performance changing point comes to lower values of data. This figure concludes that the GPU is better when the data load is high and CPU is better when data load is low. However, this is not the only factor affecting the changing point. The performance of the GPU and CPU, main memory size of CPU and GPU, complexity of the algorithm and data load are major things to consider when this critical point to be find.

However the applications developers using parallel processing of GPU should have include an option to switch processing units on their applications when the user need it because we know that the CPU is better for small inputs and small number of patterns when the GPU is better for large inputs and large number of patterns. Therefore they have to identify this critical point and should modify their algorithms to switch processing units automatically according to the previous experimental results to get better performance according to the given input.

## V. CONCLUSION

We have to do some pre-experiment before using GPU as the parallel processing unit in day to day real applications. Performance of the CPU and GPU, size of the available memory and complexity of the algorithms are the main factors to consider before publishing your applications to real world. However, average performance of machines of end users and average size of the data load of the application are very effective factors to identify the best option among these. Therefore GPU application developers have to consider these facts and modify their applications for giving effective products to clients.

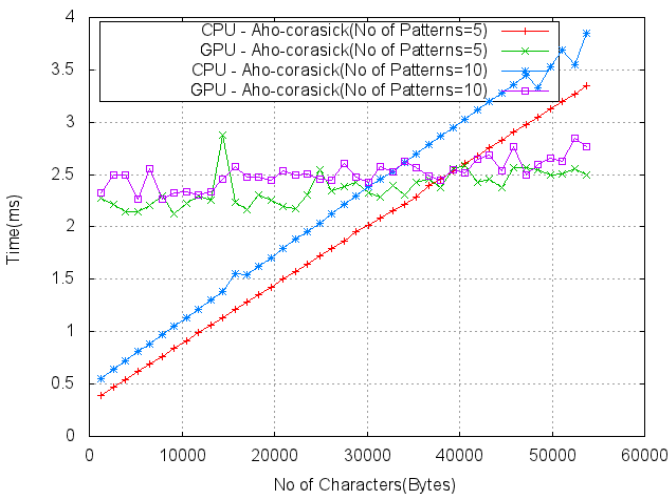

Fig. 8. Aho-Corasick vs multifast with deferent pattern sizes